\documentclass[apjl]{emulateapj}
\usepackage{upgreek}
\usepackage{newtxtext}
\usepackage{newtxmath}
\usepackage{bm,url}
\usepackage[breaklinks,colorlinks,urlcolor=blue,citecolor=blue,linkcolor=blue]{hyperref}

\begin{document}


\title{FRB\,121102 is coincident with a star forming region in its
  host galaxy} \shorttitle{FRB\,121102 is coincident with a star
  forming region in its host galaxy} \shortauthors{Bassa et al.}


\author{C.~G.~Bassa$^{1}$,
  S.~P.~Tendulkar$^{2}$,
  E.~A.~K.~Adams$^{1}$,
  N.~Maddox$^{1}$,
  S.~Bogdanov$^3$,
  G.~C.~Bower$^4$,
  S.~Burke-Spolaor$^{5,6,7}$,
  B.~J.~Butler$^5$,
  S.~Chatterjee$^8$,
  J.~M.~Cordes$^8$,
  J.~W.~T.~Hessels$^{1,9}$,
  V.~M.~Kaspi$^2$,
  C.~J.~Law$^{10}$,
  B.~Marcote$^{11}$,
  Z.~Paragi$^{11}$, 
  S.~M.~Ransom$^{12}$,
  P.~Scholz$^{13}$, 
  L.~G.~Spitler$^{14}$, 
  H.~J.~van~Langevelde$^{11,15}$
}
\affiliation{$^1$ ASTRON, the Netherlands Institute for Radio Astronomy, Postbus 2, NL-7990 AA Dwingeloo, The Netherlands}
\affiliation{$^2$ Department of Physics and McGill Space Institute, McGill University, 3600 University St., Montreal, QC H3A 2T8, Canada}
\affiliation{$^3$ Columbia Astrophysics Laboratory, Columbia University, New York, NY 10027, USA}
\affiliation{$^4$ Academia Sinica Institute of Astronomy and Astrophysics, 645 N. A'ohoku Place, Hilo, HI 96720, USA}
\affiliation{$^5$ National Radio Astronomy Observatory, Socorro, NM 87801, USA}
\affiliation{$^6$ Department of Physics and Astronomy, West Virginia University, Morgantown, WV 26506, USA}
\affiliation{$^7$ Center for Gravitational Waves and Cosmology, West Virginia University, Chestnut Ridge Research Building, Morgantown, WV 26505}
\affiliation{$^8$ Cornell Center for Astrophysics and Planetary Science and Department of Astronomy, Cornell University, Ithaca, NY 14853, USA}
\affiliation{$^9$ Anton Pannekoek Institute for Astronomy, University of Amsterdam, Science Park 904, NL-1098 XH Amsterdam, The Netherlands}
\affiliation{$^{10}$ Department of Astronomy and Radio Astronomy Lab, University of California, Berkeley, CA 94720, USA}
\affiliation{$^{11}$ Joint Institute for VLBI ERIC, Postbus 2, NL-7990 AA Dwingeloo, The Netherlands}
\affiliation{$^{12}$ National Radio Astronomy Observatory, Charlottesville, VA 22903, USA}
\affiliation{$^{13}$ National Research Council of Canada, Herzberg Astronomy and Astrophysics, Dominion Radio Astrophysical Observatory, P.O. Box 248, Penticton, BC V2A 6J9, Canada}
\affiliation{$^{14}$ Max-Planck-Institut f\"ur Radioastronomie, Auf dem H\"ugel 69, D-53121 Bonn, Germany}
\affiliation{$^{15}$ Leiden Observatory, Leiden University, P.O. Box 9513, NL-2300 RA Leiden, The Netherlands}

\begin{abstract}
  We present optical, near-\ and mid-infrared imaging of the host
  galaxy of FRB\,121102 with the Gemini North telescope, the
  \textit{Hubble Space Telescope} and the \textit{Spitzer Space
    Telescope}. The FRB\,121102 host galaxy is resolved, revealing a
  bright star forming region located in the outskirts of the
  irregular, low-metallicity dwarf galaxy. The star forming region has
  a half-light radius of 0.68\,kpc ($0\farcs20$), encompassing the
  projected location of the compact ($<0.7$\,pc), persistent radio
  source that is associated with FRB\,121102. The half-light diameter
  of the dwarf galaxy is 5 to 7\,kpc, and broadband spectral energy
  distribution fitting indicates that it has a total stellar mass of
  $M_\star\sim10^8$\,M$_\odot$. The metallicity of the host galaxy is
  low, $12+\log_{10} (\mathrm{[O/H]})=8.0\pm0.1$. The properties of
  the host galaxy of FRB\,121102 are comparable to those of extreme
  emission line galaxies, also known to host hydrogen-poor
  superluminous supernovae and long-duration $\gamma$-ray bursts. The
  projected location of FRB\,121102 within the star forming region
  supports the proposed connection of FRBs with newly born neutron
  stars or magnetars.
\end{abstract}

\keywords{galaxies: dwarf -- galaxies: star formation -- supernovae:
  general -- gamma-ray burst: general}

\section{Introduction}\label{sec:introduction}
Over a decade ago, \citet{lbm+07} presented the discovery of a
millisecond-duration radio transient whose large dispersion measure
suggested an extragalactic origin. Since then, the discovery of a
population of fast radio burst (FRB) sources
(e.g.\ \citealt{tsb+13,sch+14,mls+15}) has fueled questions pertaining
to both their physical origin and whether they can be used as probes
of the intervening (inter-)galactic material. Many theoretical models
have been proposed, including those that invoke a cataclysmic event,
and those in which repeated bursts from the same source are possible.
Regardless, the FRB phenomenon has generated great interest because
the short durations and cosmological distances necessarily imply sites
of extreme energy density.

The detection of polarization, Faraday rotation, and scintillation in
some FRBs has given clues about their origin -- in some cases
suggesting that they originate from dense and highly magnetized
environments like those of a supernova remnant \citep{mls+15}.  The
discovery of repeated radio bursts from FRB\,121102 immediately ruled
out cataclysmic progenitor models -- at least for this particular
source \citep{ssh+16a,ssh+16b}.  However, deep campaigns to search for
repeat bursts from other known FRB positions have thus far detected no
other repeaters \citep{pjk+15}, suggesting perhaps that there is more
than one class of FRB progenitor within the sample of 22 known sources
\citep{pbj+16}.

In a practical sense, the repetition of FRB\,121102 greatly
facilitates follow-up and precision localization. Monitoring of the
FRB\,121102 field using the Very Large Array in a fast-dump recording
mode \citep{lbb+15} led to a sub-arcsecond localization of the radio
bursts and the discovery of the optical host galaxy and a persistent
radio counterpart \citep{clw+17}. Very-long-baseline interferometry
(VLBI) using Arecibo and the European VLBI Network (EVN) constrained
the source of the FRB\,121102 bursts and the position of the
persistent radio source to be within 12\,mas from each other
\citep{mph+17}. This led to the conclusion that the source of the
FRB\,121102 bursts has a direct physical link to the source of the
persistent radio emission; e.g., the burst source could be embedded in
a radio-bright nebula. Spectroscopy of the optical host galaxy
identified it as a low mass, low-metallicity dwarf galaxy at redshift
$z = 0.193$ \citep{tbc+17}, constituting the first unambiguous
identification of an FRB host and its precise distance.

A tantalizing clue resulting from the identification of the host
galaxy of FRB\,121102 is that similar dwarf galaxies are known to host
long duration $\gamma$-ray bursts (LGRBs; \citealt{mkk+08}) and
hydrogen-poor superluminous supernovae (SLSN-I;
\citealt{lcb+14}). This could suggest an evolutionary link, in which
the sources of FRB events are born at the times of LGRBs and SLSN-I,
with FRBs originating from young neutron stars or magnetars resulting
from these explosions \citep{pir16,mbm17,be17,km17}.

In this paper we present multi-band optical and infrared (IR) imaging
of the host galaxy of FRB\,121102 with the Gemini North telescope, the
\textit{Hubble Space Telescope} (\textit{HST}) and the \textit{Spitzer
  Space Telescope}. In \S\,\ref{sec:observations} we report on the
observations and analysis. The morphology, environment and properties
of the host galaxy are presented in \S\,\ref{sec:results}. We discuss
our results in \S\,\ref{sec:discussion}.

\begin{figure*}
  \includegraphics[width=0.248\textwidth]{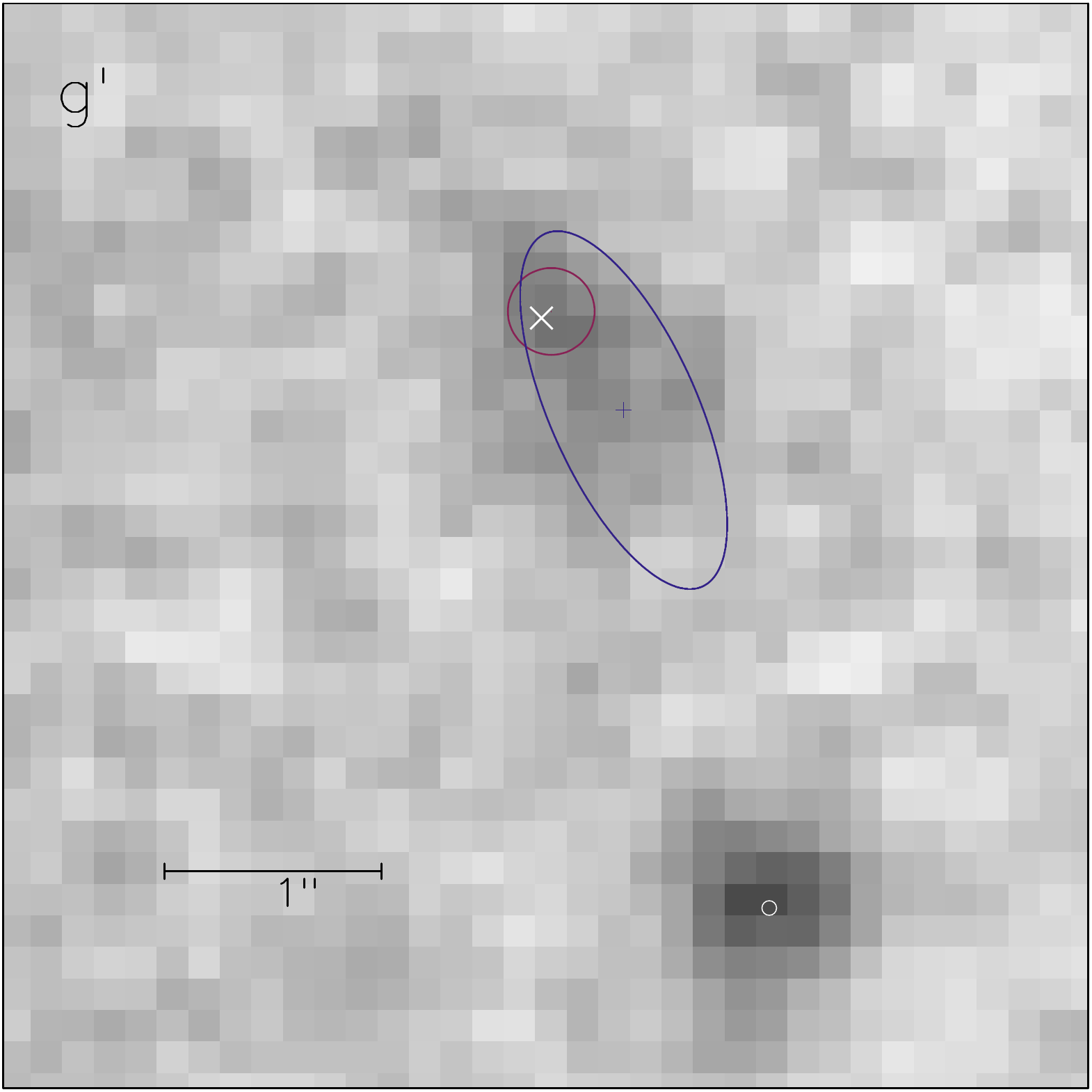}
  \includegraphics[width=0.248\textwidth]{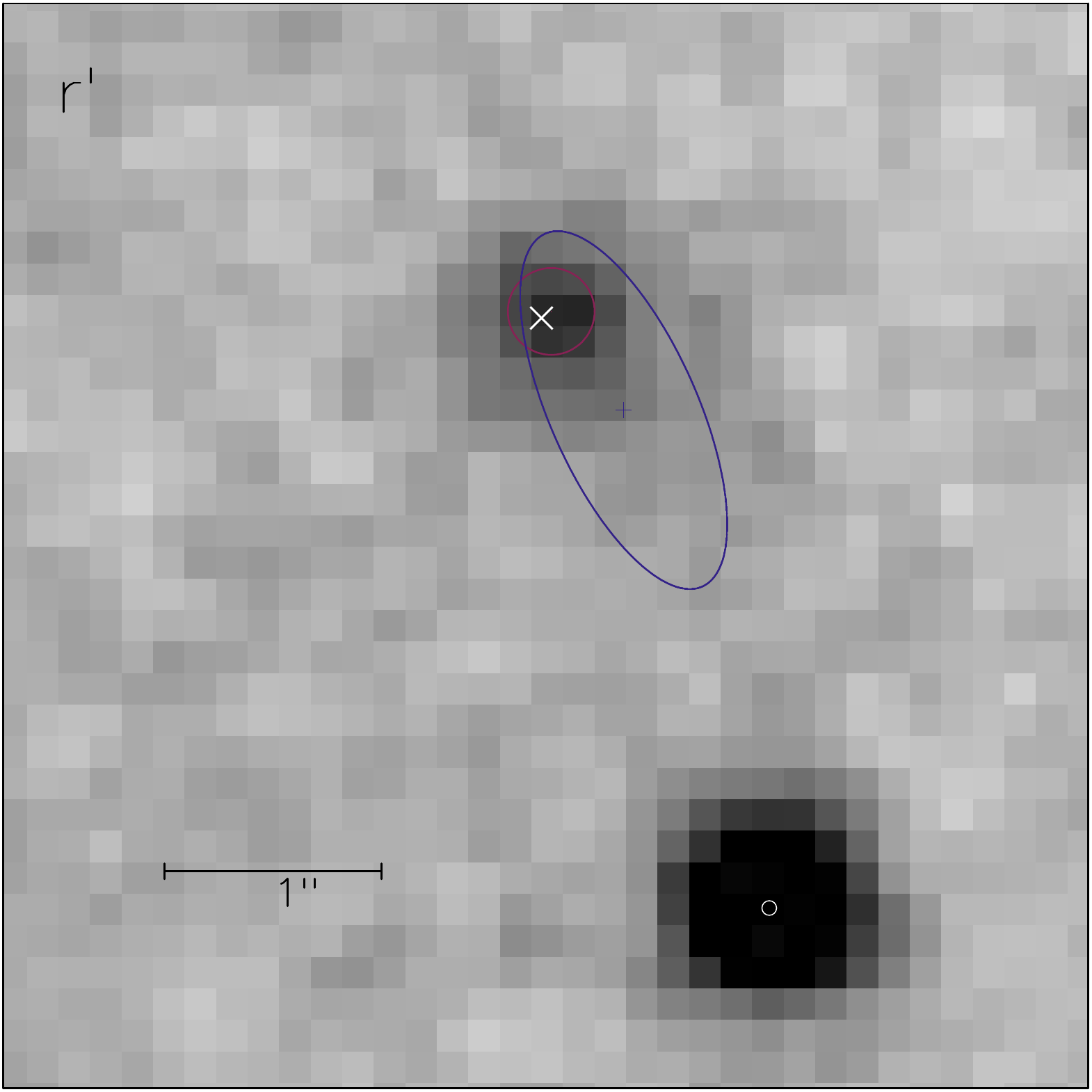}
  \includegraphics[width=0.248\textwidth]{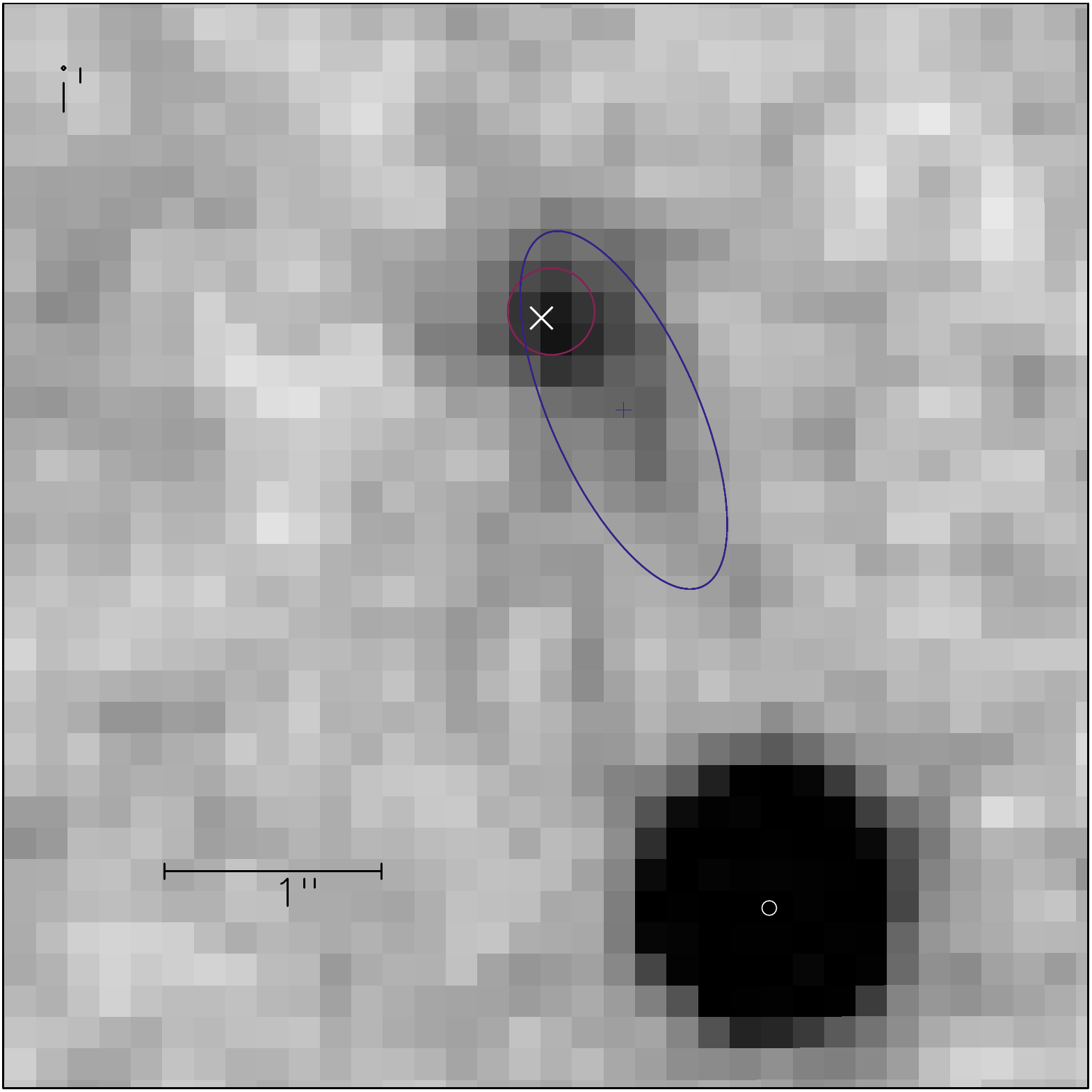}
  \includegraphics[width=0.248\textwidth]{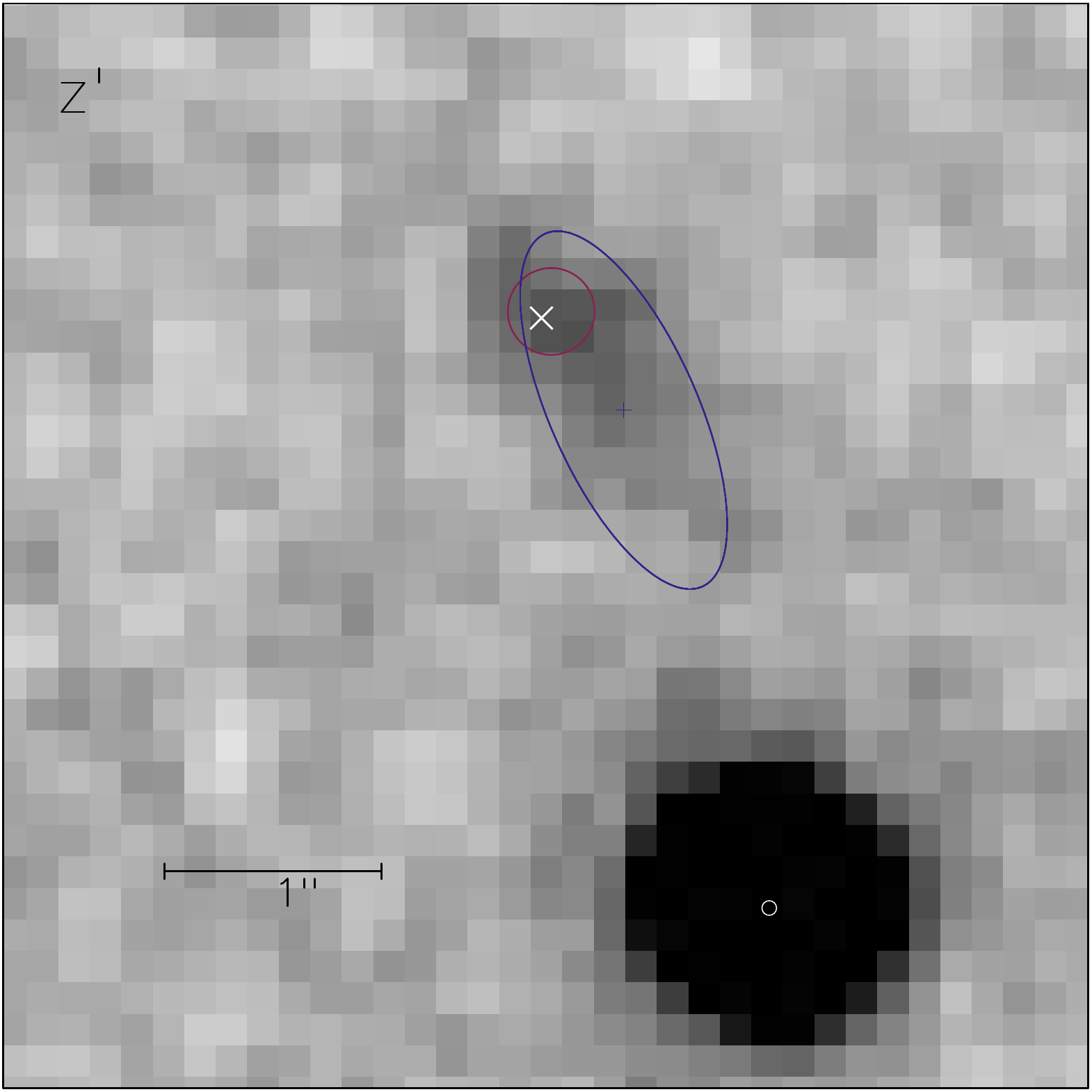}\\
  \includegraphics[width=0.248\textwidth]{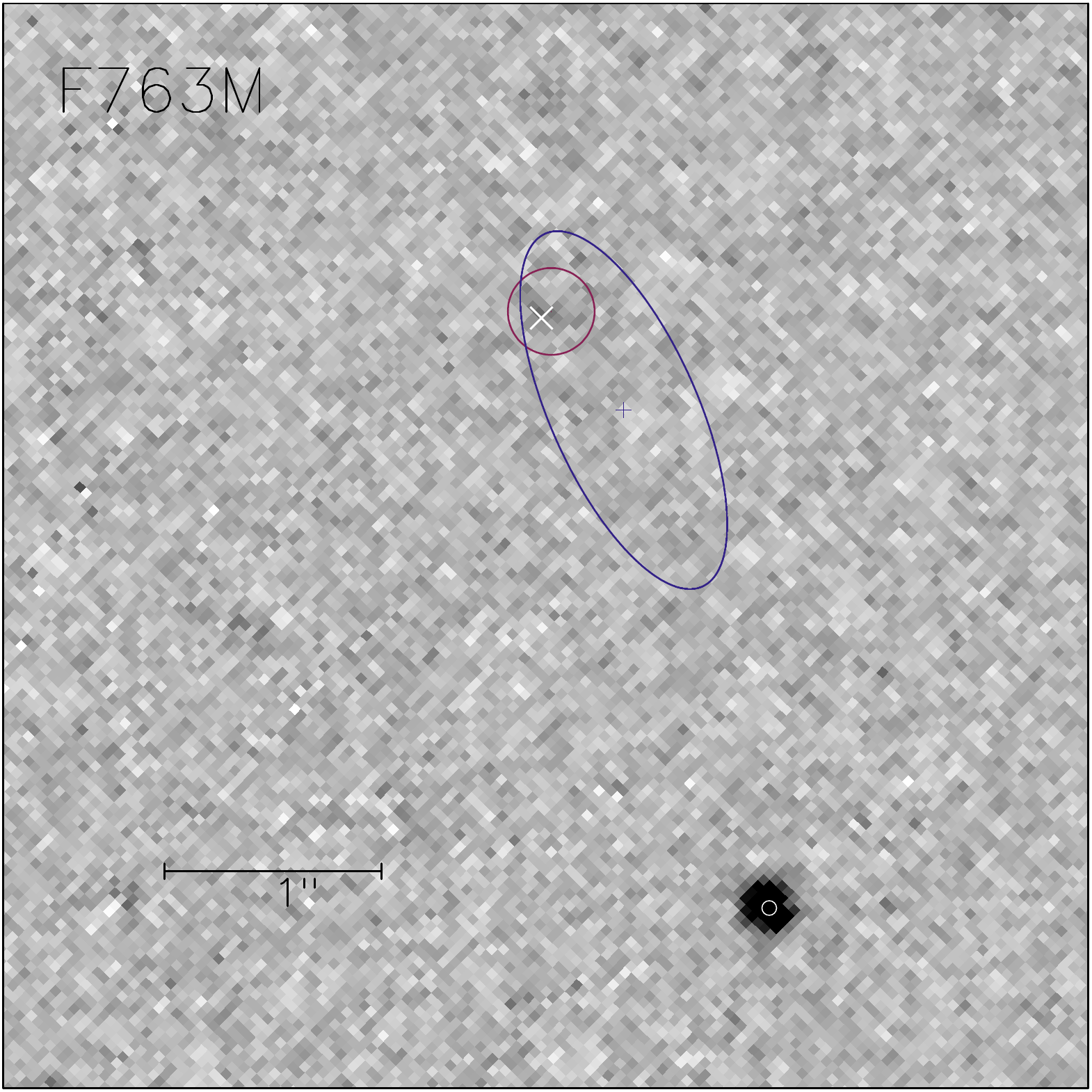}
  \includegraphics[width=0.248\textwidth]{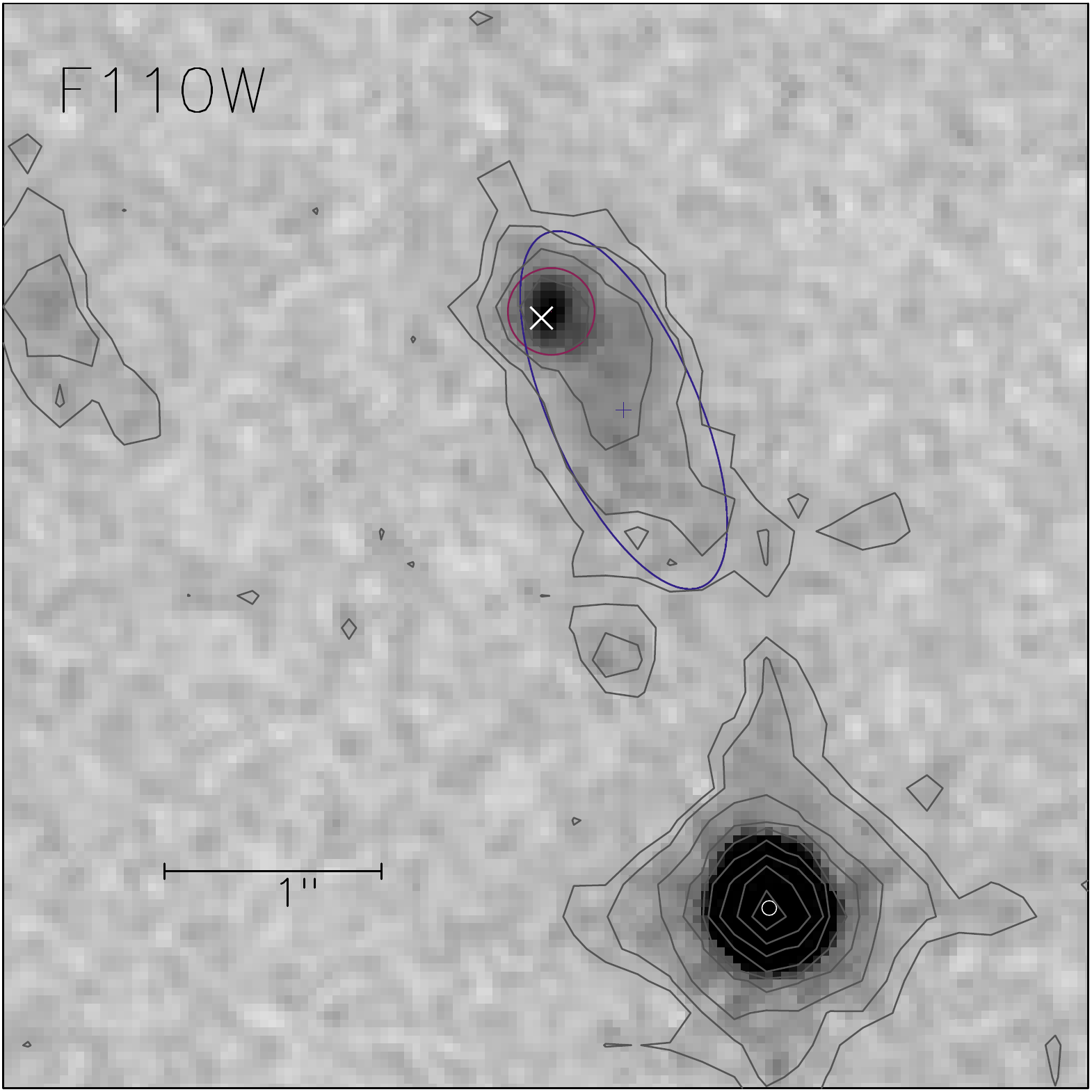}
  \includegraphics[width=0.248\textwidth]{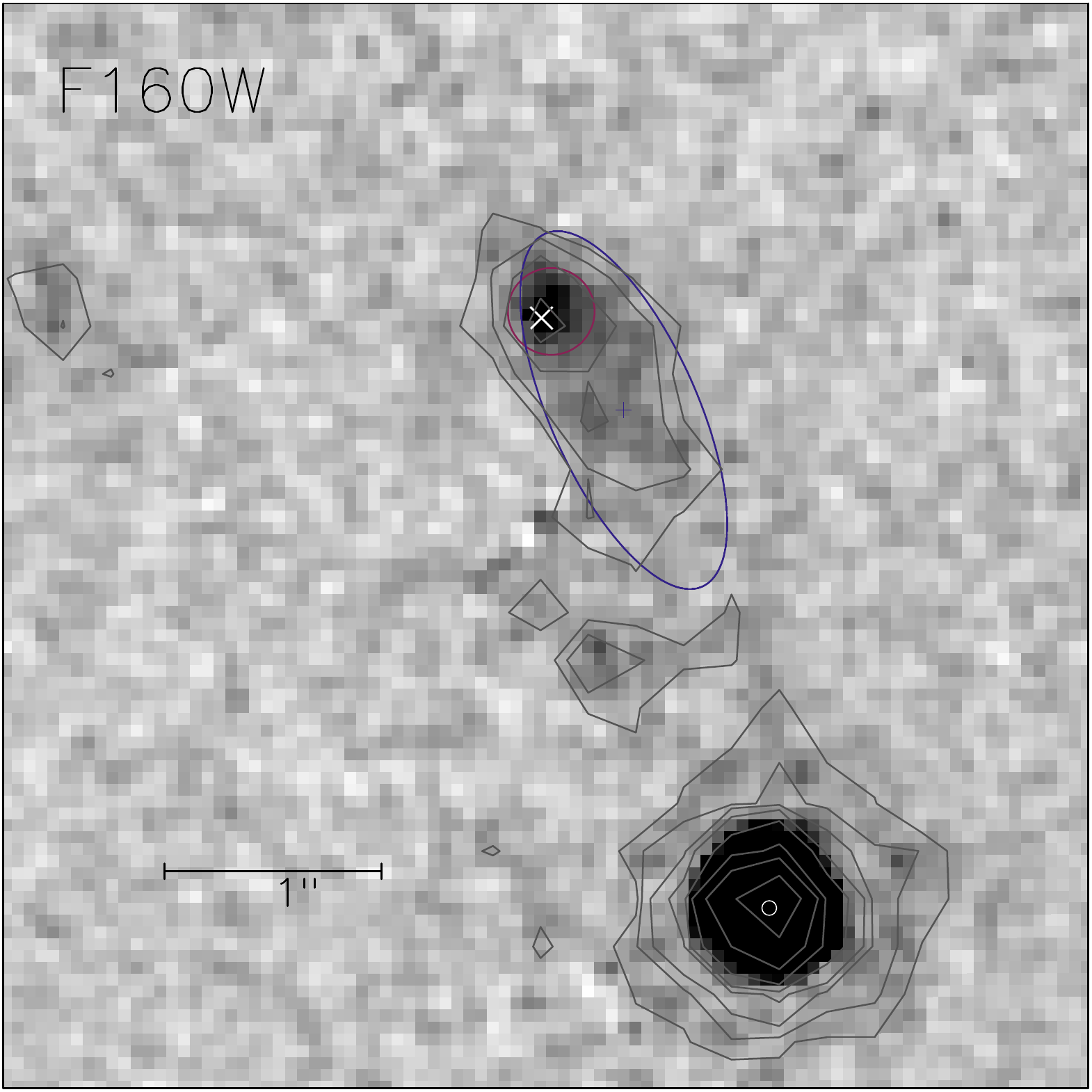}
  \includegraphics[width=0.248\textwidth]{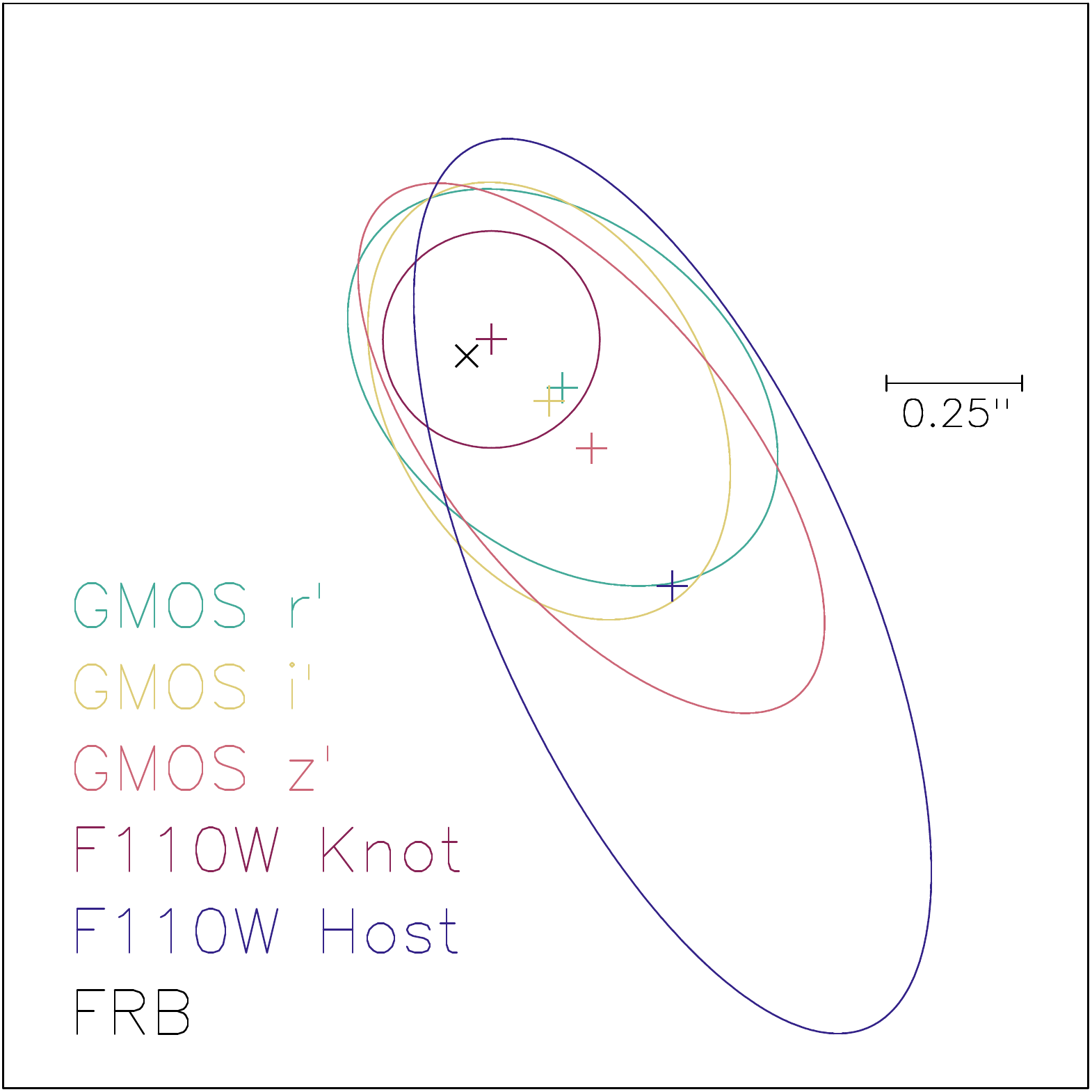}
  \caption{Subsections ($5\arcsec\times5\arcsec$) of the GMOS
    $g^\prime$, $r^\prime$, $i^\prime$, and z$^\prime$ images (top
    row), and of the \textit{HST}/WFC3 F763M (redshifted H$\alpha$),
    F110W ($J$) and F160W ($H$) images (bottom row). North is to the
    top and East to the left. The GMOS images have been smoothed with
    a Gaussian with a width of $0\farcs1$. We fit the centroid and
    half-light radius of both the bright knot and extended diffusing
    emission using the F110W image. These centroids and half-light
    radii are denoted by the red circle and the large blue ellipse,
    respectively, in each frame. Contours are overlaid on the F110W
    and F160W images to indicate the extent of the host galaxy. The
    white circle denotes the position of the reference star. The lower
    right panel ($2\arcsec\times2\arcsec$) combines the GMOS $r^\prime
    i^\prime z^\prime$ position and extent, as determined by
    \citet{tbc+17}, with those of the knot and host from the
    F110W. The black cross (white cross in the GMOS and
    \textit{HST}/WFC3 images) denotes the location of FRB\,121102 and
    the associated persistent radio source by \citet{mph+17} at 5\,GHz
    (0.6\,mas uncertainty, smaller than the symbol).  }
  \label{fig:images}
\end{figure*}

\section{Observations and analysis}\label{sec:observations}

\subsection{Hubble Space Telescope}
We acquired \textit{HST} imaging observations of the FRB\,121102 field
on 2017 February 23 using the WFC3/UVIS and WFC3/IR cameras (datasets
IDE601010, IDE601020, IDE601030 and IDE601040). We chose the F763M and
F845M filters to cover the redshifted H$\alpha$ emission line and the
continuum redward of H$\alpha$. In the near-IR we used the F110W and
F160W filters, which are comparable to the $J$ (1.1\,$\upmu$m) and $H$
(1.6\,$\upmu$m) bands. The exposure times were 1940\,s (F763M),
2560\,s (F845M), 1797\,s (F110W) and 1197\,s (F160W). We used 4-point
dither patterns appropriate for the UVIS and IR cameras to improve the
sampling of the point-spread-function (PSF) and to mitigate cosmic
rays and hot pixel issues.

The data were processed through STScI's \texttt{drizzlepac} package
\citep{ghfm12,ahc+15} to drizzle and mosaic the images. The final
resolution was chosen to include 2.5 drizzled pixels in the full-width
at half maximum (FWHM) of the PSF appropriate for each wavelength. The
images were combined with optimal weighting proportional to inverse of
the pixel variance (i.e.\ \texttt{IVM} weighting). The
\texttt{drizzlepac} pipeline also produced pixel weight images for
each filter.  Each drizzled and mosaiced image was astrometrically
matched to the \textit{Gaia} DR1 catalog \citep{bvp+16}. The
root-mean-square (rms) residuals of the astrometric calibration were
8--9\,mas for each image, using of order 50 (80) \textit{Gaia}
standards for the IR (UVIS) images.  We performed photometry on the
reduced images using the \texttt{Source Extractor} package
\citep{ba96}, generating isophotal aperture magnitudes, corrected for
blending. We used AB magnitude zero-points for each filter as defined
by
STScI\footnote{\url{http://www.stsci.edu/hst/wfc3/analysis/uvis_zpts}
  for UVIS and \url{http://www.stsci.edu/hst/wfc3/ir_phot_zpt} for
  IR}. The \texttt{Source Extractor} catalogs were then merged with
photometry in other filters.

\subsection{Spitzer Space Telescope}
We acquired \textit{Spitzer} IRAC \citep{fha+04} observations of
FRB\,121102 on 2017 January 4 with the 3.6\,$\upmu$m and 4.5\,$\upmu$m
bands (Obs ID 62322432). The observation was split into 100 dithered
exposures of 100\,s for each band. The data were processed through the
\texttt{MOPEX} software using the mosaicing and the multi-frame point
source extraction pipeline. We used the detected source catalog from
the \textit{HST} F160W image as input to the pixel response function
fitting photometry and source extraction routine. The positions of the
sources were held fixed while the fluxes were fit.

\subsection{Gemini}
Imaging observations of the host galaxy of FRB\,121102 were obtained
with the Gemini Multi-Object Spectrograph (GMOS) on the 8-m Gemini
North telescope atop Mauna Kea, Hawai'i (program GN-2016B-DD-2). On
2016 December 29 we obtained deep exposures in $g^\prime$
($12\times300$\,s) and in $r^\prime$ ($6\times250$\,s).  The
conditions during these observations were photometric, with the seeing
varying between $0\farcs6$ and $0\farcs8$. The GMOS images were read
out with $2\times2$ binning, yielding a pixel scale of
$0\farcs146$. These observations are in addition to the deep
$r^\prime$, $i^\prime$ and $z^\prime$ imaging presented in
\citet{tbc+17}, and were bias-corrected, flatfielded, registered and
co-added in an identical manner.

All co-added GMOS images were astrometrically calibrated against
\textit{Gaia} standards, using 40 to 50 unblended stars and yielding
rms residuals of 9--10\,mas.  Instrumental magnitudes of objects on
the co-added GMOS images were determined through isophotal aperture
photometry with \texttt{Source Extractor} \citep{ba96}. The
instrumental magnitudes were calibrated directly to the AB system with
photometry from Pan-STARRS\,1 DR1 \citep{cmm+16,msf+16}. Due to the
similarity between the GMOS and Pan-STARRS filters no color terms
were required, and we fitted only zero-point offsets.

\section{Results}\label{sec:results}
The drizzled WFC3 images obtained in the F110W and F160W bands show
that the object identified by \citet{clw+17} and \citet{tbc+17} as the
host galaxy of FRB\,121102 consists of a compact and bright knot,
offset from diffuse emission (Fig.\,\ref{fig:images}). We obtain
isophotal aperture magnitudes of the host galaxy (both the knot and
diffuse emission) of $m_\mathrm{F110W}=23.675(12)$ and
$m_\mathrm{F160W}=23.31(3)$ in the AB system. In the lower resolution
GMOS images these two components are blended, yielding isophotal
aperture magnitudes (AB system) of $g^\prime=25.85(12)$,
$r^\prime=25.46(14)$, $i^\prime=24.75(9)$ and $z^\prime=24.30(13)$.
In the \textit{Spitzer} images, the host galaxy flux was measured to
be $1.03\pm0.19$\,$\upmu$Jy and $<0.9$\,$\upmu$Jy ($6\sigma$) in the
3.6\,$\upmu$m and 4.5\,$\upmu$m bands, respectively.

Our $r^\prime$ and $i^\prime$-band magnitudes for the host galaxy of
FRB\,121102 are fainter by 0.4 and 0.9\,mag compared to what we
obtained in \citet{tbc+17}. We attribute this difference to an error
in the aperture photometry reported in \citet{tbc+17}. As a result of
the overestimated brightness of the host, the WFC3 UVIS exposure times
were underestimated, unfortunately leading to low signal-to-noise
ratios in the F763M and F845M images. Only the knot is detected at
$5\sigma$ in the F763M image.  We note though that the flux scale of
the GMOS spectrum of the FRB\,121102 host galaxy of \citet{tbc+17} is
is correct, as this was scaled to the $i^\prime$-band magnitude of the
reference star, which was not affected by the error in the aperture
photometry.

The high spatial resolution of the F110W and F160W images allows us to
confirm our hypothesis we suggested in \citet{tbc+17}; we identify the
knot as a star forming region responsible for the observed emission
line spectrum, while the diffuse emission corresponds to the
underlying stellar population of the FRB\,121102 host galaxy. The flux
from the star forming knot in the $r^\prime$ and $i^\prime$ bands is
dominated by the bright emission lines of H$\alpha$, H$\beta$,
[\ion{O}{3}] $\lambda4959$ and [\ion{O}{3}] $\lambda5007$. As a
result, the position and extent of the host galaxy, as we determined
in \citet{tbc+17}, reflects the differing contributions of the star
forming knot compared to the rest of the host galaxy. The emission
from the knot in the F110W and F160W bands is expected to contain
emission lines of [\ion{S}{3}] 0.907$\upmu$m, [\ion{S}{3}]
0.953$\upmu$m, \ion{He}{1} 1.083$\upmu$m, Pa$\delta$, Pa$\gamma$ and
Pa$\beta$ (\citealt{mrd+13}, see also Fig.\,\ref{fig:sed}), explaining
its brightness compared to the diffuse emission.

\subsection{Morphology}
We determine the position and extent of the star forming complex (the
knot) and the underlying stellar population in the drizzled F110W
image by modelling and jointly fitting them as two-dimensional
Gaussian or Moffat \citep{mof69} profiles. We find that the
ellipticity of the knot is close to unity, so we fit it with a
circular Moffat function instead. The knot has a radius of
$\sigma=0\farcs24(1)$, significantly larger than the radius of the
stellar PSF, for which a Moffat fit yields radii of
$\sigma=0\farcs165$. The diffuse emission appears irregular in the
near-IR images (Fig.\,\ref{fig:images}). For simplicity, we fit the
stellar population with a Gaussian profile, which yields a semi-major
axis of $\sigma_a=0\farcs66(3)$ with $b/a=0.40(2)$, and a position
angle of 66\degr. Transferring the position of the knot and the
diffuse emission to the F160W image and keeping the positions fixed,
we find comparable results for the position and size of the star
forming region. The diffuse emission prefers a larger semi-major axis
of $\sigma_a=0\farcs85(3)$ and a smaller ellipticity $b/a=0.36(2)$.

We estimate the intrinsic radius of the knot as the quadratic
difference of the observed radius and that of the stellar PSF. The
resulting half-light radius corresponds to a half-width at half
maximum (Gaussian HWHM, $1.1774\sigma$) of $0\farcs20(1)$. At a
redshift of $z=0.193$, an angle of $1\arcsec$ corresponds to a
projected distance of 3.31\,kpc, hence the half-light radius is
$0.68(3)$\,kpc. Under the assumption that the diffuse emission due to
the underlying stellar population can be represented by a Gaussian
(which, given the irregular nature of dwarf galaxies may not be
valid), the knot is located $0\farcs57$ (1.9\,kpc) from the nominal
centroid of the diffuse emission, which itself has a half-light
diameter (Gaussian FWHM) between $1\farcs5$ to $2\farcs0$ (5 to
7\,kpc). The higher spatial resolution and greater depth of the
\textit{HST} observations improve upon the $\lesssim4$\,kpc diameter
we estimated in \citet{tbc+17}. The centroid of the knot, as measured
in the drizzled F110W image, is located at
$\alpha_\mathrm{J2000}=05^\mathrm{h}31^\mathrm{m}58\fs6980(8)$,
$\delta_\mathrm{J2000}=+33\degr08\arcmin52\farcs671(10)$. This
position is consistent with that determined from the $5\sigma$
detection of the knot in the F763M image, as well as the F160W
image. The milliarcsecond-precision location of the persistent radio
source at 5\,GHz with the EVN \citep{mph+17} is offset from the center
of the star forming region by $0\farcs055(14)$ or $0.18(5)$\,kpc, but
located within the nominal half-light radius of the star forming
region. The nearby reference star is located at
$\alpha_\mathrm{J2000}=05^\mathrm{h}31^\mathrm{m}58\fs6180(8)$,
$\delta_\mathrm{J2000}=+33\degr08\arcmin49\farcs831(1)$.

\subsection{Spectral energy distribution fitting}\label{ssec:sed}
\begin{figure}
  \includegraphics[width=\columnwidth]{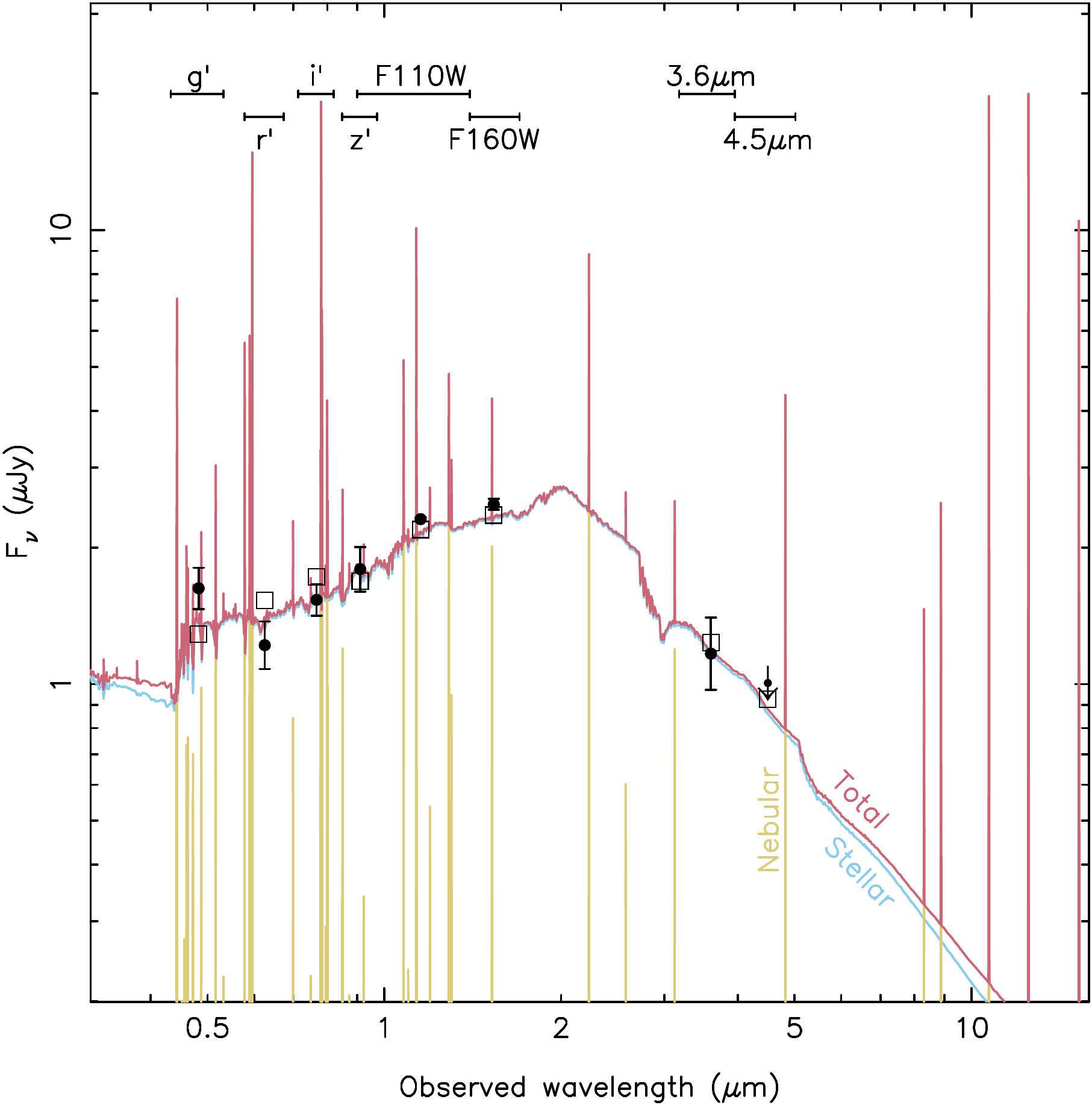}
  \caption{Spectral energy distribution (SED) for the host galaxy of
    FRB\,121102. Photometric measurements or upper limits from Gemini,
    \textit{HST} and \textit{Spitzer} are indicated with black dots
    or downward arrow, with the respective bandpasses indicated at the
    top. The stellar and nebular (from star formation) components
    making up the total emission of the SED fit are shown in different
    colors. The predicted model fluxes are shown with open squares.}
  \label{fig:sed}
\end{figure}

We use our multi-wavelength photometry to model the spectral energy
distribution (SED) of the host galaxy of FRB\,121102 with the
\texttt{CIGALE}\footnote{Available at \url{http://cigale.lam.fr/}.}
software \citep{nbg+09,sat+11}. We fit an underlying older stellar
population with a recent burst of star
formation. Figure\,\ref{fig:sed} shows the resulting SED. One of the
largest sources of uncertainties is the foreground Galactic
extinction. We used both the \citet{sfd98} and \citet{sf11} foreground
extinction values and found no appreciable differences in our
best-fitting SED and derived parameters, and we report results using
the \citet{sfd98} Galactic extinction correction. We find that the
host has negligible internal dust extinction, as expected for a
metal-poor dwarf galaxy. Our best value for the recent star formation
(averaged over the last 10 Myrs) is $0.13(4)$\,M$_\odot$\,yr$^{-1}$.
This value does not fully capture the current star formation as
measured directly from the H$\alpha$ luminosity
(0.23\,M$_\odot$\,yr$^{-1}$ with no internal extinction correction),
but is roughly consistent. We find a stellar mass of
$M_\star=(1.3\pm0.4)\times10^8$\,M$_\odot$, which corresponds to a
mass-to-light ratio of $\sim 0.6$ in the $r^\prime$-band. The stellar
mass is dominated by the older stellar population and so is
insensitive to the exact details of the recent star
formation. However, we note that the inherent uncertainties in
determining the stellar mass from SED fitting are at least a factor of
two \citep{pmt12}. The stellar mass is somewhat larger, but consistent
with, the $\sim(4-7)\times10^7$\,M$_\odot$ we estimated in
\citet{tbc+17} from the H$\alpha$ derived star formation rate.

\subsection{Metallicity}
In \citet{tbc+17} we placed rough constraints on the metallicity of
the host galaxy using the measured emission lines.  These constraints
suffered from the fact that [\ion{O}{2}] $\lambda3727$ and
[\ion{O}{3}] $\lambda4363$ were outside our wavelength coverage. Here
we use the \texttt{HII-CHI-mistry}\footnote{Available at
  \url{http://www.iaa.es/~epm/HII-CHI-mistry.html}} software
\citep{per14} to obtain a more robust constraint on the metallicity of
the host galaxy. This software uses grids of photoionization models to
derive abundances consistent with the direct method, or $T_e$ method,
wherein the electron temperature of the gas is constrained by
measuring the ratio between the [\ion{O}{3}] $\lambda4363$ and
[\ion{O}{3}] $\lambda5007$ lines. By assuming empirical laws between
$\mathrm{O/H}$, $\mathrm{N/O}$, and the ionization parameter $\log U$,
it provides reasonable metallicities even when the temperature
sensitive [\ion{O}{3}] $\lambda4363$ line is not present, as is the
case for our spectrum.  Using either the $3\sigma$ upper limit for
[\ion{N}{2}]\,$\lambda6584$ or a tentative $\sim1.7\sigma$ detection,
we find $12+\log_{10} (\mathrm{[O/H]})=8.0\pm0.1$.  While there are
inherent uncertainties in the metallicity of the host galaxy without
the measurement of more lines, we can confirm it is a low-metallicity
galaxy.

\subsection{Star formation rate}
From the Galactic extinction-corrected H$\alpha$ emission line, we
estimate the star formation rate (SFR) to be
$0.23-0.4$\,M$_\odot$\,yr$^{-1}$, with the lower value uncorrected for
internal extinction and the upper value with a correction applied
\citep{tbc+17}. Due to the negligible dust content of low-metallicity
dwarf galaxies, we expect the internal extinction to be low, and thus
we adopt the uncorrected H$\alpha$ SFR of 0.23\,M$_\odot$\,yr$^{-1}$
as the best estimate. This value is also more in line with the lower
SFR derived from the SED fitting of 0.12\,M$_\odot$\,yr$^{-1}$.

This star formation will also be observable in the radio continuum,
which has the advantage of being unaffected by dust obscuration from
the Galactic plane. Using the relation between SFR and 1.4\,GHz radio
luminosity from \citet{mcs+11}, the H$\alpha$-derived SFR corresponds
to $3$\,$\upmu$Jy. From \citet{clw+17}, the VLA 1.4\,GHz flux is
$250\pm39$\,$\upmu$Jy; however, this includes flux not only from star
formation, but also from the persistent radio source which is embedded
within the star forming region. The \textit{HST} imaging shows the
star forming region confined to a region of $\sim0\farcs2$ in radius,
scales that are resolved out in the EVN observations presented in
\citet{mph+17}. These EVN observations show that the persistent radio
source flux at 1.7\,GHz varies between 168 and 220\,$\upmu$Jy, with
flux calibration uncertainties of the order of 20 per cent. The
difference between the VLA and EVN fluxes, (i.e.\ the flux association
with star formation), is consistent with the estimated excess of a few
$\upmu$Jy.

The star forming knot is barely resolved in the \textit{HST} images,
so we have no detailed information about its internal structure. We
expect the star forming complex to be composed of a number of
individual, unresolved, star-forming regions. Giant molecular clouds
found in the LMC \citep{hwo+10} and interacting Antennae galaxies
\citep{zfb+14} reach 100\,pc or more in size. A few of such GMCs in
close proximity to each other would be sufficient to produce such a
star-forming region.

\subsection{Environment of the FRB\,121102 host galaxy}
We have made a crude estimate of the environment of the FRB host
galaxy, using the \textit{HST} F110W ($J$-band) image. First we count
the number of objects within a fixed-radius aperture of $15\farcs5$,
corresponding to a radius of 50\,kpc at the redshift of the host
galaxy, centered on the location of the FRB. This is large enough to
encompass the immediate and extended environment of the FRB host
galaxy.

As galaxy clustering strength is a function of galaxy mass, with more
massive galaxies being more strongly clustered than low-mass galaxies
\citep{zbf+02}, we select other objects in the image within
$\pm0.15$\,mag of the FRB host magnitude, but outside the FRB host
aperture. At these faint magnitudes, the objects will either be dwarf
galaxies like the FRB host (albeit at unknown redshifts), or very high
redshift, more massive galaxies. We place the same $15\farcs5$
aperture around each of these magnitude-matched galaxies and count the
number of objects, to use as control regions. No attempt to remove
foreground stars has been made, as it is assumed to be qualitatively
the same for every aperture and will thus cancel out.

For the \textit{HST} $J$-band image, there are 30 objects within the
$15\farcs5$ aperture to $J\leq26.5$ for both the FRB and 31 control
regions, but their distribution in radius is different. The FRB region
is more centrally concentrated, with 18 objects within $10\arcsec$
compared to a median value of $14\pm4$ in the control regions. Thus,
on small ($<50$\,kpc) scales, there is a marginal hint of an
over-density centered on the FRB host galaxy. On larger scales, the
environment is typical of other objects with the same apparent
magnitude. The irregular morphology of the host galaxy is consistent
with past or ongoing interactions.

\section{Discussion and conclusions}\label{sec:discussion}
We present optical, near-, and mid-IR imaging which improves on the
ground based imaging presented in \citet{tbc+17} and resolves the host
galaxy of FRB\,121102, revealing a bright knot of star formation
located in the outskirts of an irregular (5 to 7\,kpc half-light
diameter), low-mass ($M_\star\sim10^8$\,M$_\odot$), low-metallicity
($12+\log_{10}([\mathrm{O/H}])=8.0$) dwarf galaxy. The persistent
radio source that is coincident with the FRB \citep{clw+17,mph+17}, is
located within the 0.68\,kpc half-light radius of the star forming
region. Recent high spatial resolution H$\alpha$ observations confirm
the coincidence of FRB\,121102 with the star forming region in its
host galaxy, and yield similar estimates of the size of the star
forming region and the offset of the FRB with its centroid
\citep{kms+17}.

The separation of the underlying stellar population of the FRB\,121102
host galaxy from the star forming region allows us to update our
estimate of the dispersion measure (DM) that can be attributed to the
H$\alpha$ emitting gas. Following our derivation in \S\,4.3 of
\citet{tbc+17}, the smaller size of the star forming region increases
the H$\alpha$ surface density by a factor 3.3. As a result, the
maximum host DM depth increases by a factor 1.8 to
$\mathrm{DM}_\mathrm{host}\lesssim589$\,pc\,cm$^{-3} 
L_\mathrm{kpc}^{1/2}$ (with $L_\mathrm{FRB}\leqslant L_\mathrm{kpc}$),
consistent with the estimate derived by \citet{kms+17}.

As we already noted in \citet{tbc+17}, with its low mass, low
metallicity, strong emission lines and morphology of an underlying
disturbed stellar population dominated by a compact star-forming
region, the host galaxy of FRB\,121102 shows many similarities with a
class of star-forming galaxies known as extreme emission line galaxies
(EELGs; \citealt{ass+11,wsr+11}). The one possible contrasting metric
is that we find the FRB host to be in a region of average or slightly
high density, whereas EELGs are generally found in underdensities
\citep{apc+15}.  

The EELG galaxy classification encompasses several sub-classes of
galaxies found in the literature. At the low-mass end of the EELG
population are compact and ultracompact blue dwarf galaxies (CBD,
UCBD; \citealt{tm81,cvc06}). \textit{HST} imaging of a sample of very
low redshift UCBDs reveals remarkable similarities to the FRB host
galaxy, showing objects with diffuse, irregular structure, punctuated
by compact regions of intense star formation spanning tens to hundreds
of parsecs, offset from the galaxy center (\citealt{cvc06}). We note
that the specific SFR (sSFR; SFR divided by stellar mass) and
metallicity of the FRB host are comparable to those of high redshift
($z\sim5$) Lyman break galaxies (LBGs; \citealt{gsdl16}). This
indicates that galaxies similar to the repeating FRB host were common
in the early universe, so we might expect similar FRBs at higher
redshifts.

Of particular interest are the hydrogen-poor superluminous supernovae
(SLSN-I), which preferably occur in EELGs \citep{lsk+15,pqy+16}. There
is some disagreement as to whether the SLSN-I hosts are also drawn
from the same galaxy pool as the hosts of long-duration $\gamma$-ray
bursts (LGRBs), with some authors finding the hosts similar
\citep{lcb+14} and others finding differences \citep{alp+16}. The
locations of SLSN-I are found to trace the UV light of their host
galaxies \citep{lcb+15}, though less so than LGRBs, which strongly
prefer the bright, inner regions of their hosts \citep{bbf16}. The
coincidence of FRB\,121102 with the star formation complex, which is
the brightest part of its host galaxy, further strengthens the
resemblance between FRBs and SLSN-I/LGRBs. This observed coincidence
is supported by models of magnetar birth \citep{kb10,woo10}, which are
believed to be born in the collapse of massive stars. This could
suggest an evolutionary link, in which neutron stars or magnetars are
born as LGRB or SLSN-I and evolve into FRB-emitting sources
\citep{pir16,mbm17,be17,km17,okm17}. Volumetric rate estimates of
star-forming dwarf galaxies, along with SLSN-I, LGRB and FRB rates by
\citet{nwb+17} and Law et al.\,(2017, submitted), indicate broad
consistency. The coincidence of FRB\,121102 with a star forming region
--- in a host galaxy that is similar in type to those preferentially
hosting SLSN-I and LGRBs --- suggests that targeted searches for radio
bursts or compact persistent radio counterparts, similar to that of
FRB\,121102, can be a valuable new approach to complement ongoing,
wide-field searches for FRBs. The discovery of even a single FRB
source in such a targeted search would greatly strengthen the
evolutionary connection already suggested by FRB\,121102.

\acknowledgments Support for \textit{HST} program GO-14890 was
provided by NASA through a grant from the Space Telescope Science
Institute, which is operated by the Association of Universities for
Research in Astronomy, Inc., under NASA contract NAS\,5-26555. This
work is based on observations obtained at the Gemini Observatory
(GN-2016B-DD-2), which is operated by the Association of Universities
for Research in Astronomy, Inc., under a cooperative agreement with
the NSF on behalf of the Gemini partnership: the National Science
Foundation (United States), the National Research Council (Canada),
CONICYT (Chile), Ministerio de Ciencia, Tecnolog\'{i}a e
Innovaci\'{o}n Productiva (Argentina), and Minist\'{e}rio da
Ci\^{e}ncia, Tecnologia e Inova\c{c}\~{a}o (Brazil).  This work is
also based on observations made with the Spitzer Space Telescope,
which is operated by the Jet Propulsion Laboratory, California
Institute of Technology under a contract with NASA.

CGB and JWTH acknowledge support from the European Research Council
under the European Union's Seventh Framework Programme (FP/2007-2013)
/ ERC Grant Agreement nr.\ 337062 (DRAGNET; PI Hessels). SPT
acknowledges support from the McGill Astrophysics Fellowship. EAKA is
supported by TOP1EW.14.105, which is financed by the Netherlands
Organisation for Scientific Research (NWO). JWTH acknowledges funding
from an NWO Vidi fellowship. LGS gratefully acknowledges support from
the ERC Starting Grant BEACON under contract no. 279702 and the Max
Planck Society. BM acknowledges support by the Spanish Ministerio de
Econom\'ia y Competitividad (MINECO) under grants
AYA2016-76012-C3-1-P, and MDM-2014-0369 of ICCUB (Unidad de Excelencia
`Mar\'ia de Maeztu'). This work has been supported by NSF award
\#1458952.


\bibliographystyle{aasjournal}

\end{document}